\documentstyle[psfig]{elsart}
\def\be{\begin{eqnarray}}
\def\ee{\end{eqnarray}}

\begin{document}
\begin{frontmatter}

\title{COMPUTING TECHNIQUES FOR TWO-LOOP CORRECTIONS
TO ANOMALOUS MAGNETIC MOMENTS OF LEPTONS}

\author{Andrzej Czarnecki and Bernd Krause}
\address{Institut f\"ur Theoretische Teilchenphysik,
 Universit\"at Karlsruhe,
D-76128 Karlsruhe, Germany }
\thanks{e-mail: ac@ttpux8.physik.uni-karlsruhe.de,
bk@ttpux8.physik.uni-karlsruhe.de}

\begin{abstract}
Two-loop electroweak  corrections to the electron, muon,
and tau lepton anomalous magnetic moments have been
computed recently.  Effects of hadronic contributions to the photon
propagator in two-loop QED corrections have also been reanalysed.  The
common technique used in both calculations was asymptotic expansion of
Feynman diagrams in the ratio of lepton mass to some heavy mass
scale.  In this talk we present some details of this method paying
particular attention to the expansion of multi-scale diagrams.  
\end{abstract} 


\end{frontmatter}
 
\section{Introduction}    

Anomalous magnetic moments of leptons $a_l\equiv (g_l-2)/2$ have been
providing stimulus for the development of quantum field theory as well as
its precise test ever since the first measurement of $a_e$ in 1947
 \cite{Kusch}.  Later precise measurements of $a_e$ have helped to
establish quantum electrodynamics as one of the best tested physical
theories.  Since the relative contribution of heavy fields to $a_l$
scales like $m_l^2$, muons are more sensitive to hadronic ($a_\mu^{\rm
had}$) or electroweak ($a_\mu^{\rm EW}$) effects (and to possible new
physics) than electrons.  An upcoming experiment
E821 \cite{Hughes92} at Brookhaven National Laboratory is going to be
the 
first one to test the electroweak loop effects in $a_\mu$.  Because of
the high precision of this experiment a full calculation of two-loop
electroweak effects was carried out \cite{CKM95,CKM96}.  It was found
that the two-loop effects reduce $a_\mu^{\rm EW}$ by 22.6\% from $195
\times 10^{-11}$ to $151(4) \times 10^{-11}$.

In the present talk we present some details of those calculations.  
The method of asymptotic expansions will be discussed and explained with
an example of   a non-trivial two-loop diagram with three mass
scales.  The same method can also be applied to the determination of the
kernel functions for the hadronic effects in the photon propagator;
their application to the calculation of $a_l^{\rm had}$ will be presented.

\section{Asymptotic Expansions with Multiple Mass Scales}
For the two-loop Feynman diagrams with more than one mass scale there exist
practically no analytical results.  Therefore, a calculation of
two-loop electroweak corrections to $a_l$ can only be done using one
of two methods:
numerical integration or analytical expansion in the small parameters,
like the ratio of the lepton and the weak boson masses. The latter
approach has several advantages: analytical results can be obtained to
any order in the small mass ratio (in particular we can get an exact
coefficient of the large logarithms); divergent diagrams are easily
treated; 
and the summation over the large number of diagrams
does not induce rounding errors present in the numerical approach.
The expansion has to be done carefully
even if a diagram contains widely separated mass
scales: the integration over
virtual momenta and the expansion of the integrand do not in general
commute.  A rigorous procedure for these so called asymptotic expansions
was developed \cite{Smi94}.
Below we describe an application of this formalism to
problems with more than one expansion parameter, i.e.~with three or
more different masses or external momenta.

\subsection*{Example of a 3-scale-problem}

We consider as a practical example a contribution to muon's $g-2$ in
which the external 
photon couples to a closed $\tau$-loop, which in turn is connected to
the muon line by a Z boson and a photon (fig.~(1); we omit the
external photon
coupling in the figure.)

We are interested in an expansion in the muon and the $\tau$ masses.  
In the first step, we consider all masses and momenta as small
compared to $M_Z$.
Following the rules of the Heavy Mass Expansion (HME) we have to sum
over subgraphs $\gamma$ of the initial diagram $\Gamma$ such that each
$\gamma$  contains all the lines with the large mass and 
consists of connectivity components which are 1 PI w.r.t. lines with
small masses \cite{Smi94}.

This rather formal statement is easy to understand with the help of
fig.~(1).  In each subgraph we  expand the propagators in
small masses and momenta (in our example $m_\tau$, $m_\mu$ and the
external momentum $q$, $q^2=m_\mu^2$.)
A co-subgraph (the remainder of the original diagram in which a
subgraph has been contracted to a point) has to be evaluated exactly.
For instance, the first subgraph in fig.~(1) is the initial two-loop
diagram, expanded in $m_\tau$, $m_\mu$ and $q$.  In the following two
subgraphs one $\tau$ line forms a co-subgraph.

After the first step, the heavy mass scale $M_Z$ has been integrated out.
The last co-subgraph, however, still contains two different scales
$m_\tau$ and $m_\mu$.  
In order to deal with it, we now consider $m_\tau$ as the
large mass scale and apply the HME to this co-subgraph (second line in
fig.~(1)).
Iterative application of the HME along
these lines allows a consistent expansion of a given diagram in terms
of powers and logarithms of the masses and momenta, as long as there
is a clear hierarchy between the scales. We end up with
one- or two-loop single scale integrals which can be evaluated
analytically. 

In the calculation of $a_l$ we encounter diagrams with both $W$ and
$Z$ bosons; their masses are similar.
In such case we may proceed in the
following way:  first expand in the difference of the
two similar scales. 
The number of scales is now reduced by one, and we may
continue with the HME. 

\section{Parametrization of 
two-loop electroweak contributions to $g-2$ of the
electron, muon and $\tau$ lepton}

Including two-loop corrections, 
the electroweak contribution to the anomalous magnetic
moment of a lepton with mass $m_l$ can be parametrized as a product of
the one-loop expression
and a correction factor $\left( 1+ C{\alpha\over \pi} \right)$

\begin{eqnarray}
a_l^{\rm EW} = a_l^{\rm 1-loop}
\left( 1+ C{\alpha\over \pi} \right)\:,\qquad
a_l^{\rm 1-loop} \approx
{5\over 3}{G_\mu m_l^2\over 8\sqrt{2}\pi^2}
\end{eqnarray}
It is natural to separate the subset of the two loop electroweak
contributions which contain a closed fermion loop:
\begin{eqnarray}
C= C^{\rm ferm} + C^{\rm bos} \; .
\end{eqnarray}
A calculation of the two-loop electroweak contributions to $a_\mu$ was
presented in some detail in \cite{CKM95,CKM96,CaKr96}.  The results for
$a_e$ and $a_\tau$ were presented in ref. \cite{CKM96}.  Below we
discuss in some detail the connection between $a_{e,\tau}$ and $a_\mu$.

\subsection*{Bosonic Contributions}

In ref. \cite{CKM96}
it was found that due to accidental cancellations the bosonic
contribution to $a_\mu$ is very
well approximated by terms containing large logarithms:
\be
C_\mu^{\rm bos}({\rm 2-loop})
\approx \ln {M_W^2\over m_\mu^2}\cdot
\left(-{13\over 3}+{92\over 15}s_W^2-{184\over 15}s_W^4 \right)\:.
\label{eq:reslog}
\ee

In diagrams without fermion loops the only mass scales are
the lepton mass and much larger masses of the weak
bosons. Therefore, the magnetic anomalies of $e$ and $\tau$ leptons can be
obtained from eq.(\ref{eq:reslog}) by
replacing $m_\mu$ by $m_e$ or $m_\tau$.
The numerical values are presented in table (1)\footnote{For the
entries in the table the full formula including the non-logarithmic terms was
used.}.\\

\subsection*{Fermionic Contributions}

In order to obtain the bosonic contributions to $g-2$
of the electron and the $\tau$ lepton a simple
rescaling of masses in the muon result 
could be performed.  
This cannot be done so easily in the diagrams containing a fermion
loop connected to the muon line by a photon and a $Z$ boson because
the internal fermion introduces an additional mass scale.

In the case of the $\tau$ we put 
$m_\tau=m_c$ and employ the formula for $\Delta C^{\mu}_{1d}$ in
eq. (16) of ref. \cite{CKM95} for the charm quark.
The final result for the fermionic contributions to $a_\tau$ is
\begin{eqnarray}
C_{\tau}^{\rm ferm} &=& -{9\over 5}\left(\ln {  M_Z^2\over
 m_\tau^2} +{1\over3}\ln {M_Z^2\over m_b^2} \right)
- {109\over 30} -{8\over 45}\pi^2
-{3\over16}{m_t^2\over s_W^2 M_W^2}
\\
&&
-{3\over10 s_W^2}\ln{m_t^2\over M_W^2}
-{8\over5}\ln{m_t^2\over M_Z^2}-{7\over10 s_W^2}+\Delta C_{Higgs}
\nonumber 
\end{eqnarray}

For the electron the fermionic contribution is
\begin{eqnarray}
C_{\rm electron}^{\rm ferm} &=& -{18\over 5}\ln { (m_u m_c M_Z)^{4/3}\over
(m_d m_s m_b)^{1/3}  m_e m_\mu m_\tau}
-{31\over 10} +{8\over 15}\pi^2 -{7\over10 s_W^2}\nonumber \\
&&
-{3\over16}{m_t^2\over s_W^2 M_W^2}-{3\over10 s_W^2}\ln{m_t^2\over M_W^2}
-{8\over5}\ln{m_t^2\over M_Z^2}
+\Delta C_{Higgs} 
\end{eqnarray}

\section{Higher Order Hadronic Contributions}

The main goal of the forthcoming BNL experiment \cite{Hughes92} is to
measure electroweak loop effects; at the same time, however, it will
also be sensitive to higher order hadronic effects.
These are ${\cal O}(\alpha^3)$ 
and fall into two different classes: diagrams with hadronic self
energy insertion in the photon propagator, and
light-by-light scattering diagrams. The latter will not be
discussed here. For the former, the high precision
of the BNL experiment requires a reconsideration of the
corresponding kernel functions \cite{had96}.
To this end, it is straightforward to apply the asymptotic
expansion method, as presented in the first section, also to the
analytical calculation of the higher order kernel functions for the
muon and the electron. 
The idea is to write the original three-loop diagrams with hadronic self
energy insertion in the photon propagator as a dispersion integral
over an effective photon ``mass'' $\sqrt{s}$.
Since the lower limit
of the dispersion integral is the $\pi^+\pi^-$ threshold, we may
perform an expansion of the corresponding Feynman diagram
w.r.t. ${m^2\over s}$, where $m$ is the muon or the electron mass.
In this way the problem is reduced to computing single scale diagrams;
the final integration over $s$ has to be done
numerically, including experimental data for $R(s)={\sigma(e^+e^-\to
hadrons)\over \sigma(e^+e^-\to \mu^+\mu^-)}$. Main
advantages of this approach are that
it is very simple to obtain numerically well-behaved analytical
expressions with sufficient accuracy
and the error in the numerical integration is reduced to the final
$s$-integral over data.
Along these lines the higher order hadronic contributions to $g-2$ of
leptons have been computed and previous calculations \cite{BR75} could
be checked 
by an independent method \cite{had96}. For the muon a
shift by $-11\cdot 10^{-11}$ w.r.t. previous work \cite{kno} was found --
which is somewhat smaller but still of the same order of magnitude as
the experimental precision 
of the BNL experiment. The results are summarized in table (2). 

\section{Acknowledgement}
We thank Professor William Marciano for collaboration on anomalous
magnetic moments of leptons.
This research was supported by BMBF 057KA92P and by ``Graduiertenkolleg
Elementarteilchenphysik'' at the University of Karlsruhe.

\section*{Tables}
\begin{tabular}{|c|c|c|c|c|} \hline 
Lepton\rule[-3mm]{0mm}{8mm} & $C^{\rm bos}(2-{\rm loop})$ & 
 $C^{\rm ferm}(2-{\rm loop})$ & $a^{\rm EW}_{\rm lepton}$ 
&$a^{\rm EW}_{\rm 2-loop}/a^{\rm EW}_{\rm 1-loop}$ \\ \hline\hline 
$\mu\rule[-3mm]{0mm}{8mm}$ &$-47.3$  & $-50.0$ & 
$151(4)\cdot 10^{-11}$ & $ -22.6\%$ \\ \hline
$e\rule[-3mm]{0mm}{8mm}$ & $-85.4$ & $-64.1$ & 
$3.0(1)\cdot 10^{-14}$&$ -35\%$ \\ \hline
$\tau\rule[-3mm]{0mm}{8mm}$ & $-27.1$ &$-37.8$ & 
$4.7(1)\cdot 10^{-7}$&$ -15\%$ \\ \hline
\end{tabular}\\[2mm]
Table 1:
Two-loop electroweak contributions to the anomalous magnetic
moment of the muon, electron and $\tau$ lepton, separated into the
bosonic and fermionic subsets, $C_{\rm bos}$ and  $C_{\rm ferm}$, and
the total electroweak correction $a_l = (g_l-2)/2$.
 Here, $\sin^2\theta_W = 0.224$ and $M_H = 250$ GeV 
were used; in $C^{\rm ferm}$ terms ${\cal O}(1-4\sin^2\theta)$ were
neglected.  In $C^{\rm bos}$ $\sin^2\theta$ was used as an expansion
parameter and the first four terms of the expansions were retained.
The two-loop corrections reduce the one-loop results by
the amount given in the last column.\\[2mm]

\begin{tabular}{|c|r|c|} \hline
Lepton\rule[-3mm]{0mm}{8mm} & $a_{\rm had., leading}$ (ref. \cite{Jeg96}) & 
 $a_{\rm had., higher\: order}$  \\ \hline\hline 
$\mu\rule[-3mm]{0mm}{8mm}$ &$7023.5\pm58.5\pm140.9\cdot 10^{-11}$  &
$-101\pm 6\cdot 10^{-11}$\\ \hline
$e\rule[-3mm]{0mm}{8mm}$ & $1.8847\pm0.0165\pm0.0375\cdot 10^{-12}$ &
$-2.25\pm 0.05\cdot 10^{-13}$\\ \hline
$\tau\rule[-3mm]{0mm}{8mm}$ & $338.30\pm1.97\pm9.12\cdot10^{-8}$ &
 $7.6\pm0.2\cdot10^{-8}$ \\ \hline
\end{tabular}\\[2mm]
Table 2: Higher order hadronic contributions to the $g-2$ of leptons.

\section*{Figures}

\begin{minipage}{16.cm}
\hspace*{.8cm}
\[
\mbox{
\hspace*{-30mm}
\begin{tabular}{ccccccccc}
\psfig{figure=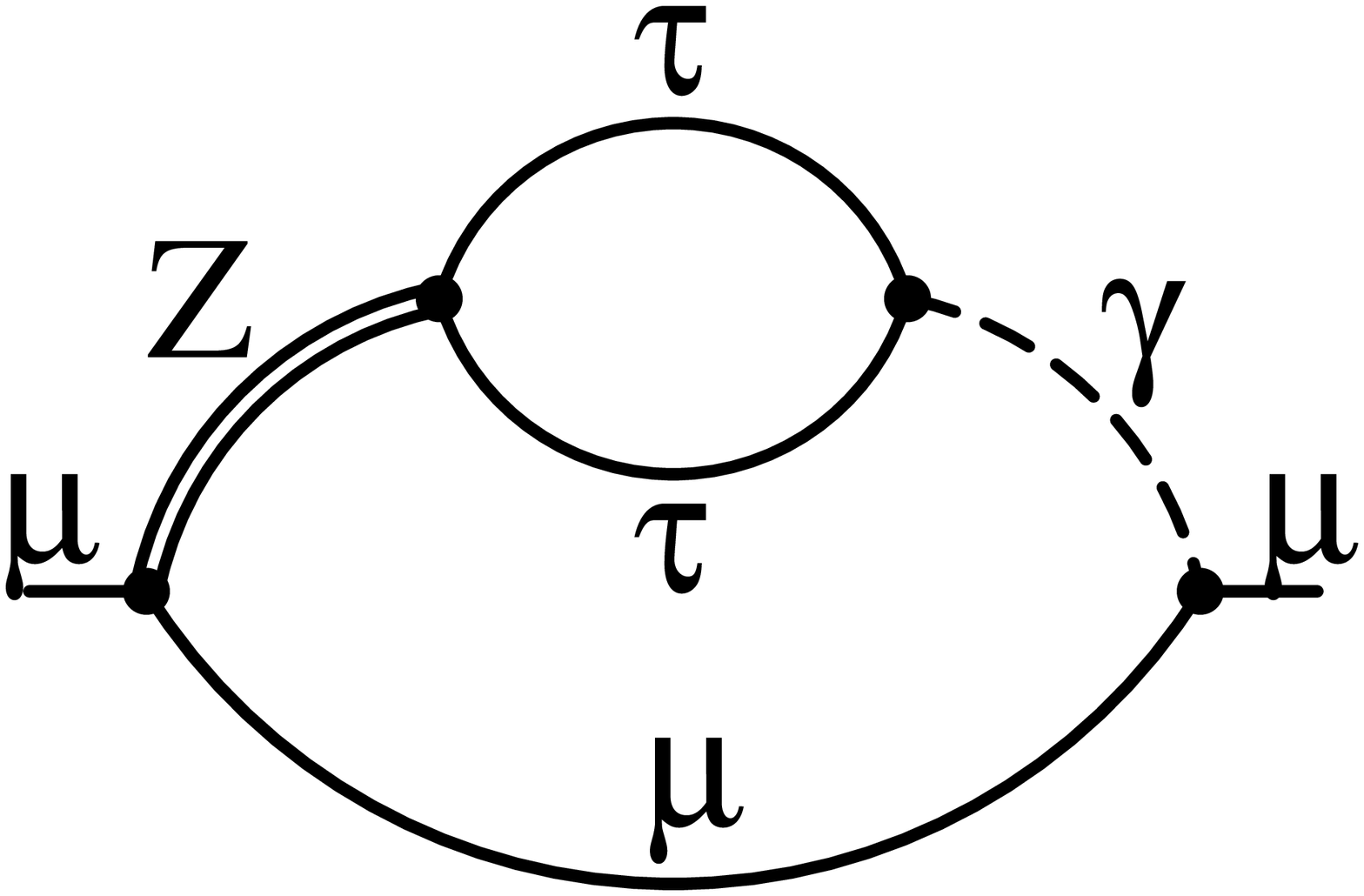,width=17mm,bbllx=210pt,%
bblly=410pt,bburx=630pt,bbury=550pt} 
\hspace*{-5mm}
&=&
\hspace*{2mm}
\psfig{figure=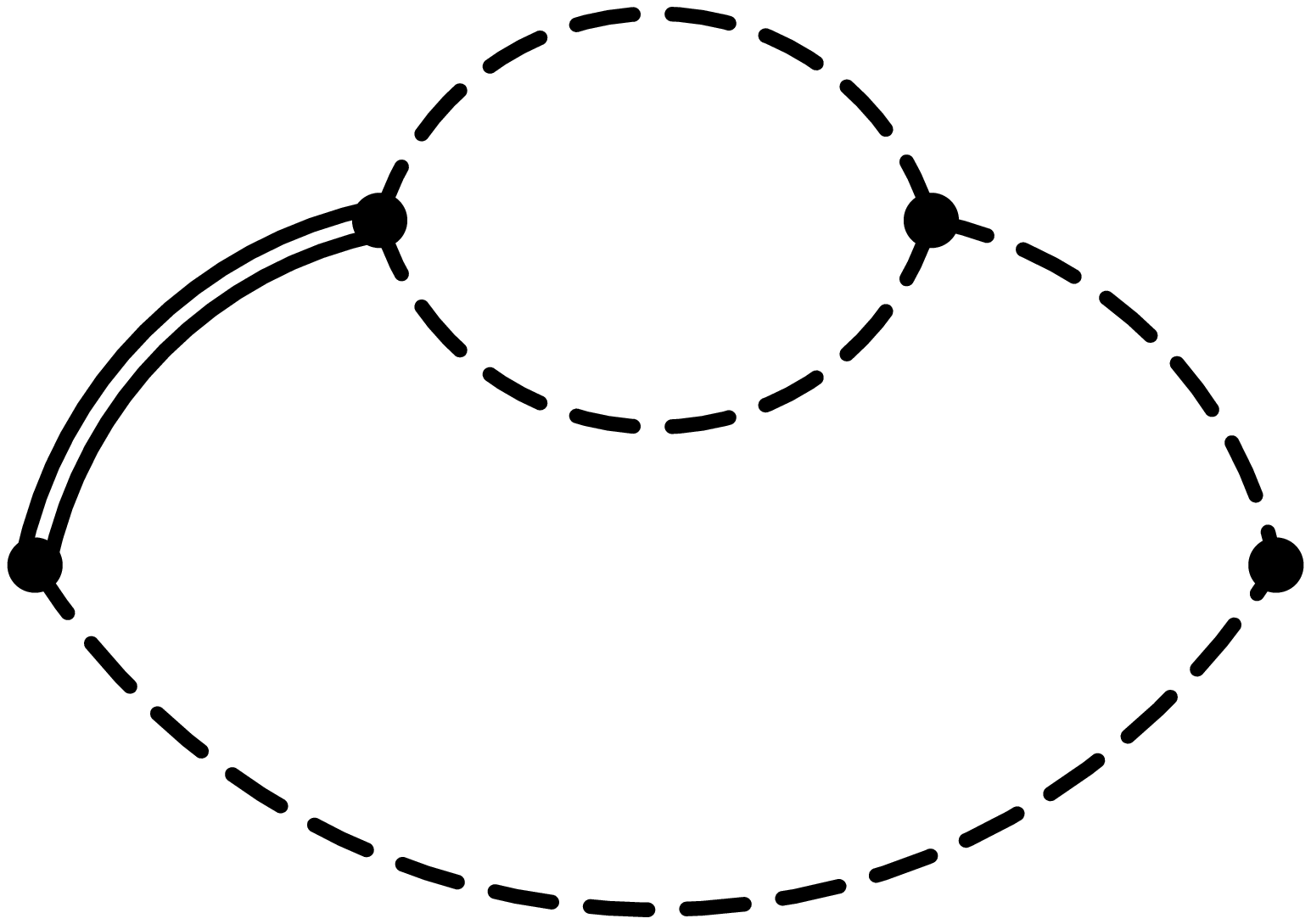,width=17mm,bbllx=210pt,%
bblly=410pt,bburx=630pt,bbury=550pt}
\hspace*{-10mm}
&\hspace*{-6mm}+\hspace*{-7mm}&
\hspace*{-6mm}
\psfig{figure=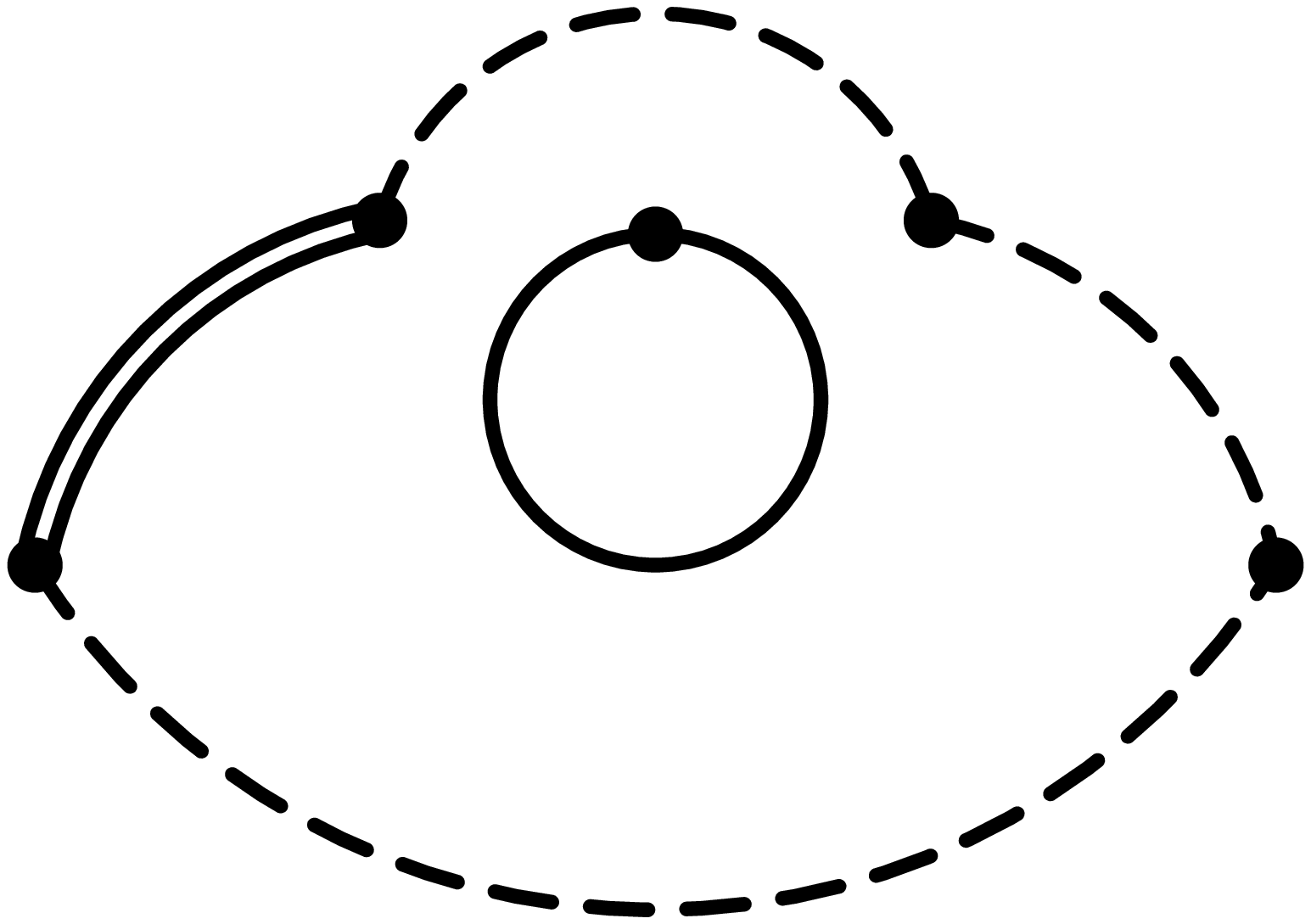,width=17mm,bbllx=210pt,%
bblly=410pt,bburx=630pt,bbury=550pt}
\hspace*{-10mm}
&\hspace*{-7mm}+\hspace*{-1mm}&
\hspace*{2mm}
\psfig{figure=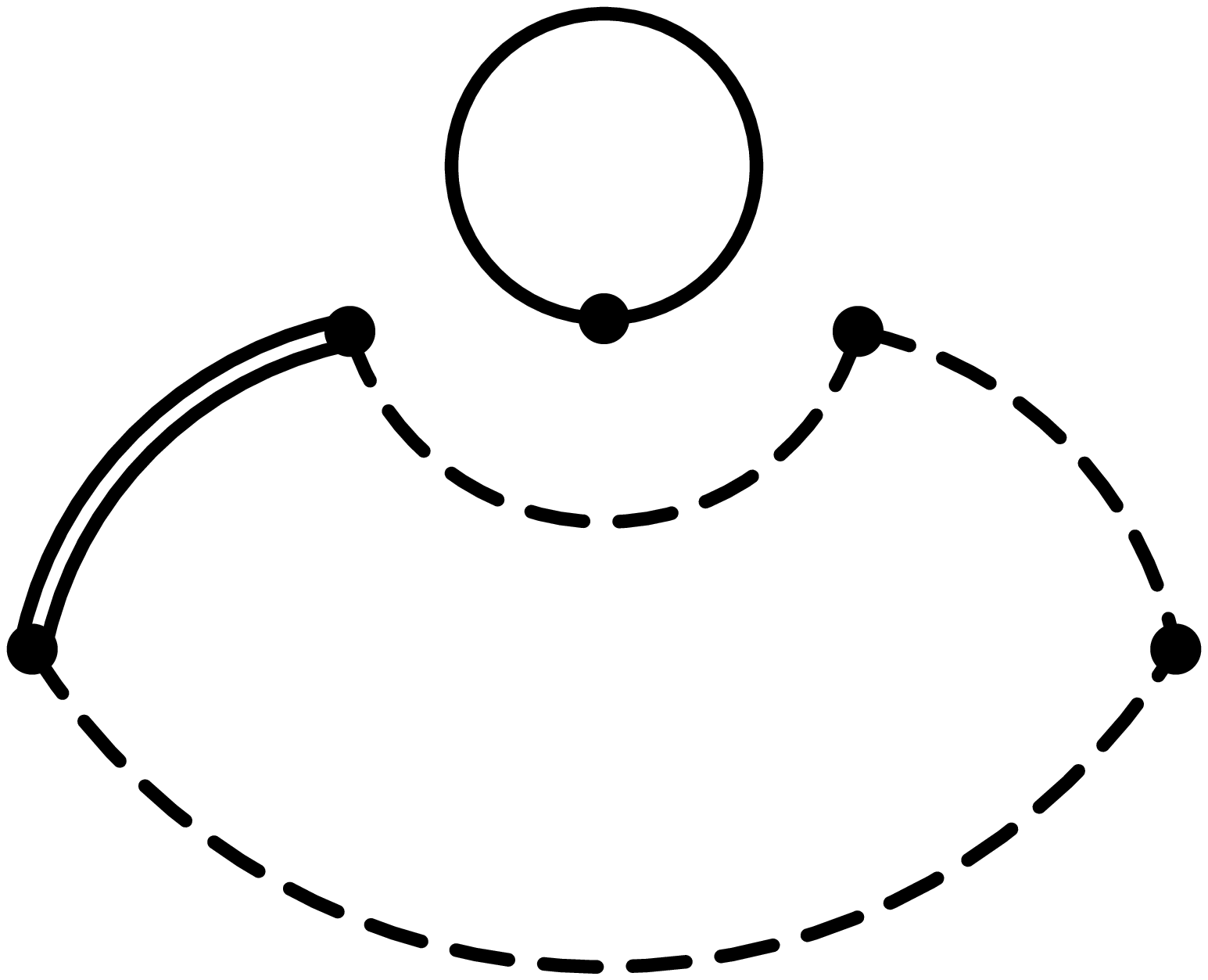,width=17mm,bbllx=210pt,%
bblly=410pt,bburx=630pt,bbury=550pt}
\hspace*{-7mm}
&+&
\hspace*{6mm}
\psfig{figure=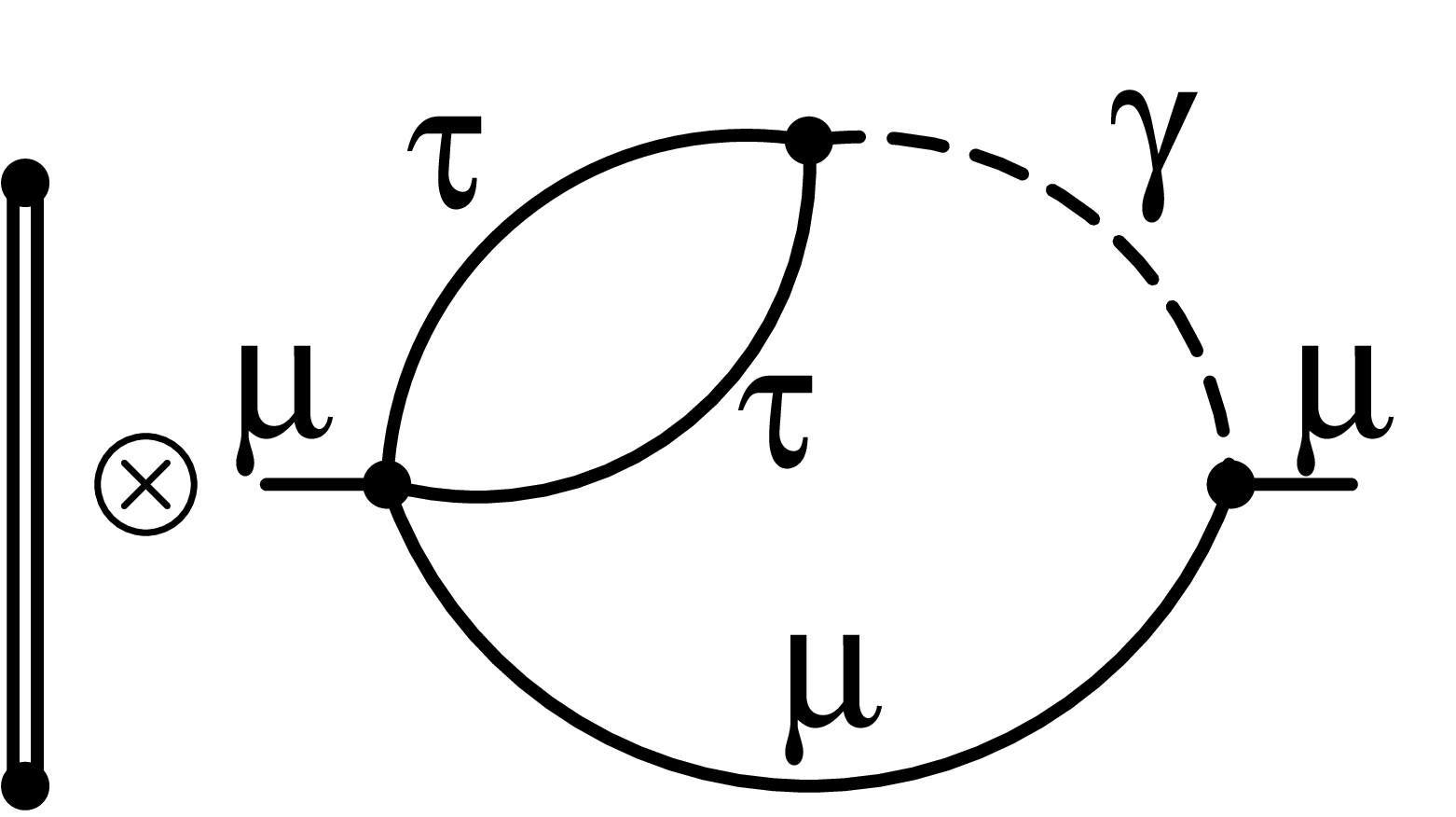,width=17mm,bbllx=210pt,%
bblly=410pt,bburx=630pt,bbury=550pt}
\\[10mm]
\psfig{figure=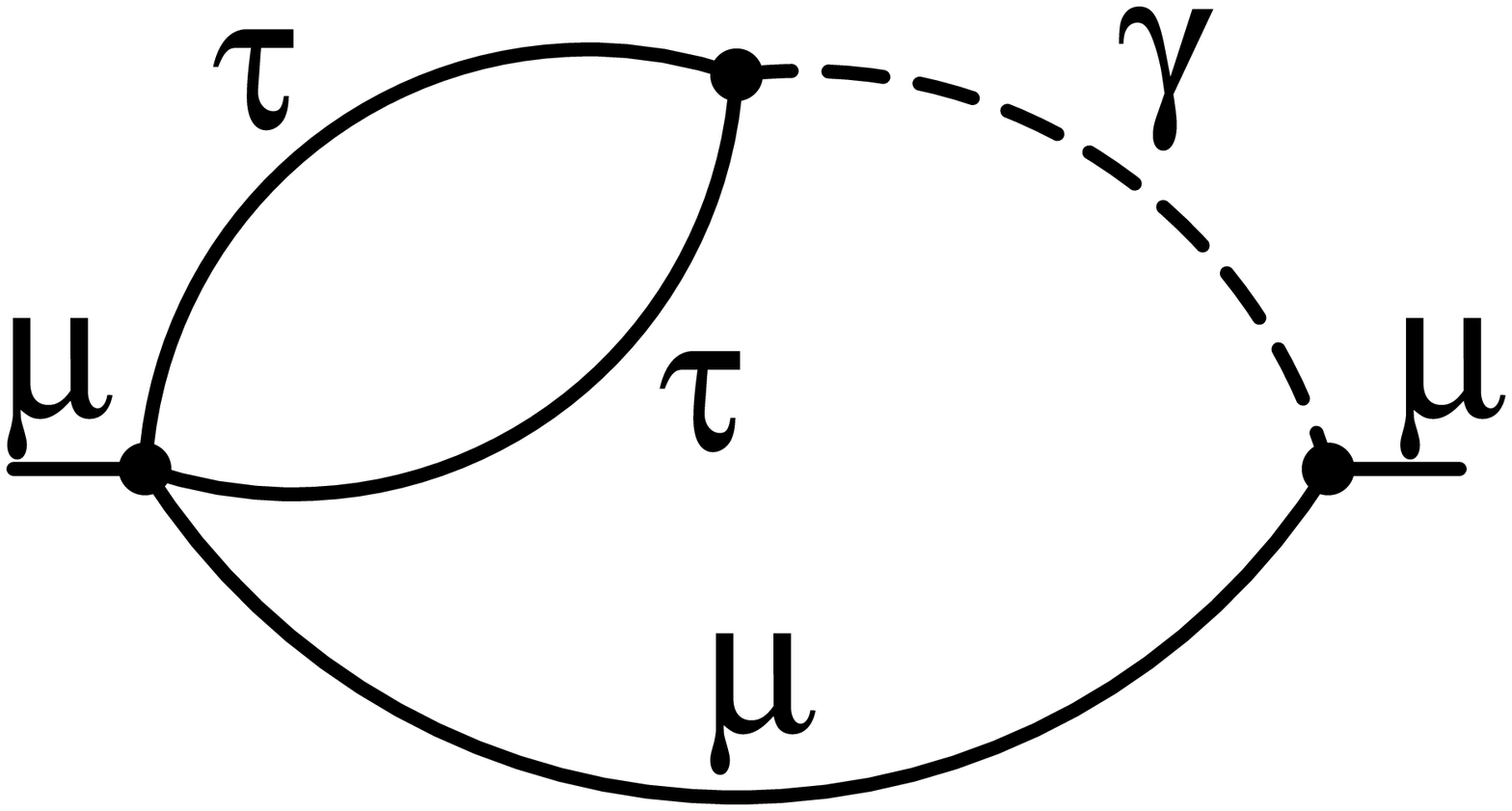,width=17mm,bbllx=210pt,%
bblly=410pt,bburx=630pt,bbury=550pt} 
\hspace*{-5mm}
&=&
\hspace*{2mm}
\psfig{figure=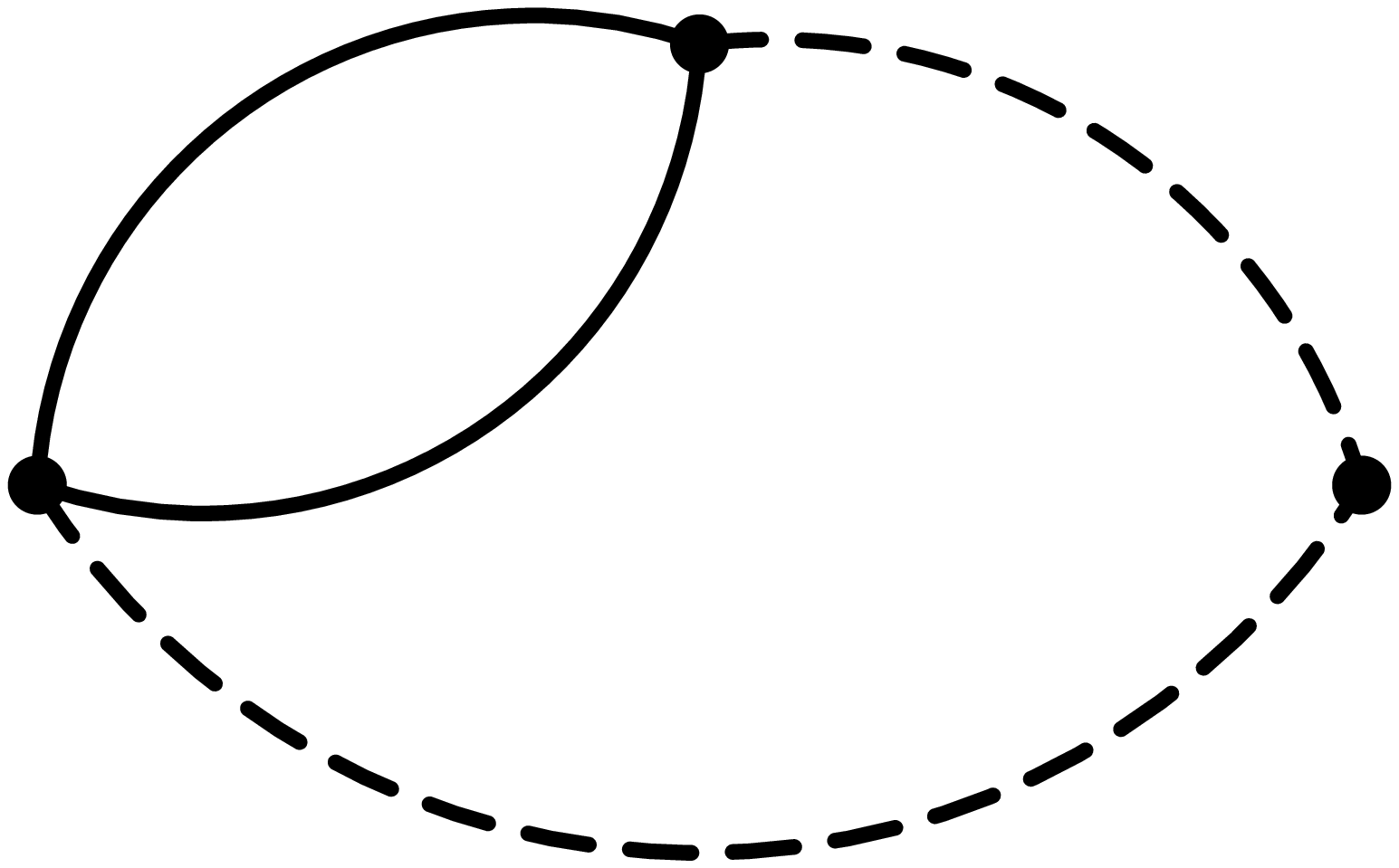,width=17mm,bbllx=210pt,%
bblly=410pt,bburx=630pt,bbury=550pt}
\hspace*{-7mm}
&+&
\hspace*{2mm}
\psfig{figure=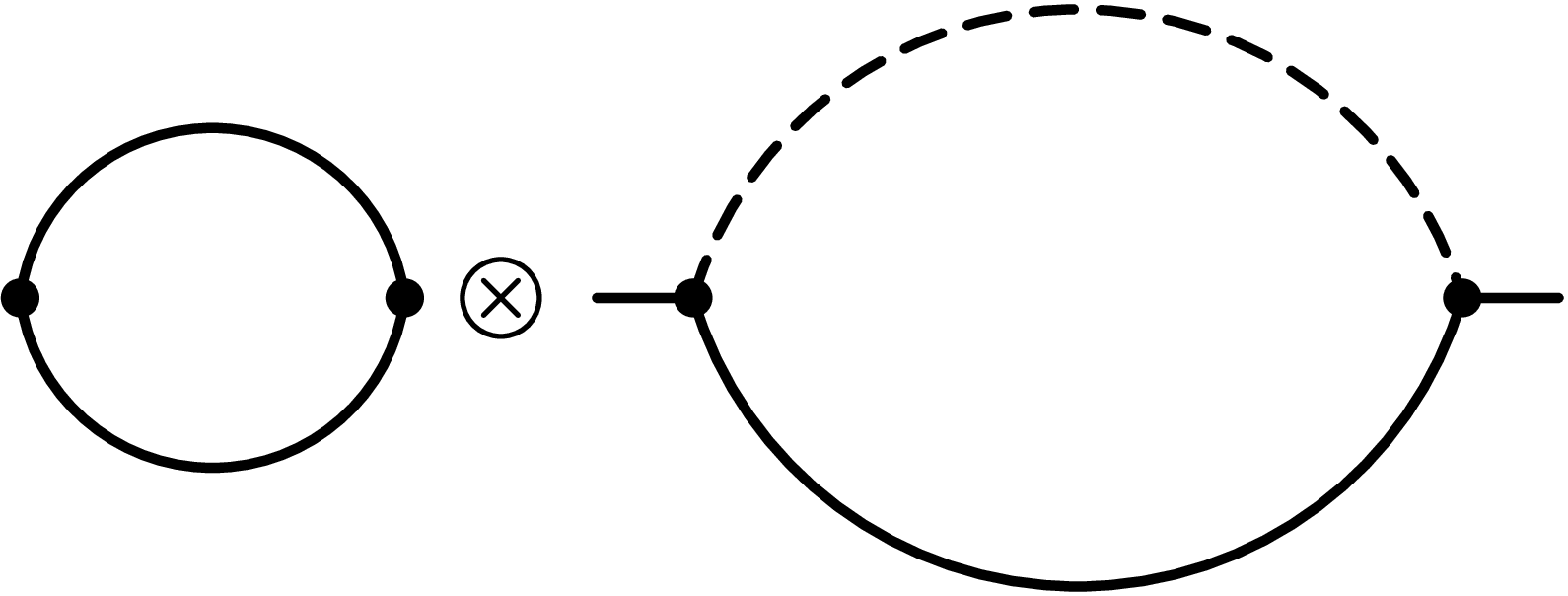,width=17mm,bbllx=210pt,%
bblly=410pt,bburx=630pt,bbury=550pt}
\\[10mm]
\end{tabular}}
\]
\end{minipage}

Figure 1: Example of the decomposition of a
three-scale-problem into single scale diagrams. Dashed lines denote
massless propagators.\\
\vspace{0.5cm}

\end{document}